\documentclass[conference, 10pt,letterpaper]{IEEEtran}
\usepackage{cite}
\usepackage{amsfonts,latexsym,amssymb}
\usepackage{amsmath,amsthm}
\usepackage{graphicx}
\usepackage{subfigure}
\usepackage{algorithm}
\usepackage{algorithmicx}
\usepackage{algpseudocode}
\usepackage{graphics}
\usepackage{epsfig}
\usepackage{cases}
\usepackage{color}
\usepackage{url}
\usepackage{bm}
\usepackage{dsfont}
\usepackage{multirow}
\usepackage{booktabs}
\floatname{algorithm}{Algorithm}
\usepackage[letterpaper, left=0.625in,right=0.625in,top=0.7in,bottom=0.97in]{geometry}
\usepackage{flushend}

\def\BibTeX{{\rm B\kern-.05em{\sc i\kern-.025em b}\kern-.08em
    T\kern-.1667em\lower.7ex\hbox{E}\kern-.125emX}}

\IEEEoverridecommandlockouts
\begin{document}
\title{\huge Personalized Federated Deep Reinforcement Learning for Heterogeneous Edge Content Caching Networks
\thanks{This work was supported in part by the Big Data Intelligence Centre of HSUHK. Tan Li and Zhen Li are equally contributing to this work. (\emph{Corresponding author: Tan Li.})}
}


\author{
    \IEEEauthorblockN{Zhen Li\IEEEauthorrefmark{1}, Tan Li\IEEEauthorrefmark{2}, Hai Liu\IEEEauthorrefmark{2}, Tse-Tin Chan\IEEEauthorrefmark{1}}
    \IEEEauthorblockA{\IEEEauthorrefmark{1} Department of Mathematics and Information Technology, The Education University of Hong Kong, Hong Kong SAR, China}
    \IEEEauthorblockA{\IEEEauthorrefmark{2} Department of Computer Science, The Hang Seng University of Hong Kong, Hong Kong SAR, China}
    \IEEEauthorblockA{E-mails: zhen\_li@ieee.org, tanli@hsu.edu.hk, hliu@hsu.edu.hk,  tsetinchan@eduhk.hk}
}

\maketitle


\begin{abstract}
Proactive caching is essential for minimizing latency and improving Quality of Experience (QoE) in multi-server edge networks. Federated Deep Reinforcement Learning (FDRL) is a promising approach for developing cache policies tailored to dynamic content requests. However, FDRL faces challenges such as an expanding caching action space due to increased content numbers and difficulty in adapting global information to heterogeneous edge environments. In this paper, we propose a Personalized Federated Deep Reinforcement Learning framework for Caching, called PF-DRL-Ca, with the aim to maximize system utility while satisfying caching capability constraints. To manage the expanding action space, we employ a new DRL algorithm, Multi-head Deep Q-Network (MH-DQN), which reshapes the action output layers of DQN into a multi-head structure where each head generates a sub-dimensional action. We next integrate the proposed MH-DQN into a personalized federated training framework, employing a layer-wise approach for training to derive a personalized model that can adapt to heterogeneous environments while exploiting the global information to accelerate learning convergence. Our extensive experimental results demonstrate the superiority of MH-DQN over traditional DRL algorithms on a single server, as well as the advantages of the personal federated training architecture compared to other frameworks.

\end{abstract}


\section{Introduction}

With the explosive growth of content volume and user demands in the era of edge computing, proactive caching has emerged as a pivotal technology for enhancing Quality of Experience (QoE) in content delivery networks \cite{cache1,cache2}. By intelligently prefetching and caching popular content at edge servers close to end users, proactive caching can significantly reduce latency and alleviate bandwidth bottlenecks on the network backbone \cite{cache3}. However, one of the key challenges in realizing the full potential of proactive caching lies in the fact that content popularity patterns are often unknown a priori and can exhibit complex spatial and temporal dynamics \cite{dynamic}.

For the proactive caching problem at individual edge servers, two mainstream approaches have been explored. The first one follows the idea that predict-first then cache \cite{predection}. This two-stage methodology first employs machine learning techniques to predict or learn user preferences or content popularity patterns and then performs cache placement based on these predictions. For instance, convolutional neural networks (CNN) \cite{cnn} and recurrent neural networks (RNN) \cite{rnn} are used for offline training of prediction models, while \cite{MAB,MAB2} adopt a multi-armed bandit (MAB) based approach for online updating. Once predictions are obtained, greedy algorithms are typically used for cache placement decisions. However, this decoupled two-stage approach suffers from issues such as error propagation and amplification across the two optimization stages, as well as the need for complex feature engineering. To circumvent these limitations, Deep Reinforcement Learning (DRL) based cache placement has recently gained traction \cite{drl, dqn, LiZ, sac, ppo}. This end-to-end approach directly learns the optimal caching policy to maximize target performance metrics, without the intermediate step of predictions. Various DRL algorithms, such as Deep Q-Network (DQN) \cite{dqn}, Soft Actor-Critic (SAC) \cite{sac}, and Proximal Policy Optimization (PPO) \cite{ppo}, have been successfully applied to the caching problem, demonstrating superior performance compared to conventional two-stage methods. 

While learning-based proactive caching methods for individual edge servers have shown promise, they face new challenges in the emerging 5G/6G edge intelligence paradigm. Learning-based algorithms often require extensive training data for the models to converge to desirable performance, which can be hindered by the limited data observed, e.g., little user feedback, at a single edge server. To overcome this data scarcity issue, collaborative learning across nodes has been explored. In particular, Federated Deep Reinforcement Learning (FDRL) based caching has attracted significant attention~\cite{fdrl1,fdrl2,fdrl3,fdrl4}, where each edge server trains a local DRL model for caching decisions using their local observation and the central cloud server periodically collects and aggregates the locally learned models during each communication round. 
However, current FDRL-based caching frameworks face two key challenges: 

1) \textit{The curse of dimensionality in the caching action space.} For a cache problem with $C$ files, there are $2^C$ possible cache configurations, making the discrete action space prohibitively large for traditional DQN \cite{dqn} to handle effectively. 
On the other hand, actor-critic methods like SAC \cite{sac} and PPO \cite{ppo}, which are designed for continuous action spaces, do not natively support discrete cache actions. To extract discrete actions from their parameterized action vectors, a sampling process is necessary, which incurs less sample efficiency and high variance.

2) \textit{Heterogeneity across edge server environments.} Existing FDRL works\cite{fdrl1,fdrl2,fdrl3,fdrl4} overlook the heterogeneity in user populations across different edge servers, which can exhibit distinct content popularity patterns due to demographic variations, necessitating personalized caching policies. A naive global model aggregation and dissemination approach fails to adapt to such heterogeneous local environments, leading to suboptimal performance. Therefore, a personalized strategy \cite{pfl} is required to unlock the full potential of federated learning for caching.

To address the aforementioned challenges of dimensionality and heterogeneity, we propose a Personalized FDRL Caching (PF-DRL-Ca) framework, aiming to maximize the system utility by jointly considering user satisfaction and cache placement costs. In our framework, each edge server trains a DRL model to handle high-dimensional action spaces using locally collected user response data. The local models are then uploaded to a central cloud server for personalized aggregation, allowing each edge server to extract global knowledge while adapting to the unique characteristics of its local user requests. The key contributions of this work are summarized as follows.

$\bullet$ For each edge server in our PF-DRL-Ca, we propose a novel DRL framework, called Multi-head Deep Q-Network (MH-DQN), to manage the exploding growth in the number of possible cache actions. In this model, each head of the network output layer generates a one-dimensional action, transforming the number of output spaces from \(2^C\) to \(2C\), where $C$ is the total number of contents.

$\bullet$ Treating each MH-DQN-aided edge server as a participant, we introduce a personalized federated training framework for collaborative learning. Specifically, each edge server uploads only a subset of base layers to a central server for aggregation. After that, it combines local personalized layers and globally aggregated base layers to generate a personalized model that is adaptable to heterogeneous environments.

$\bullet$ We first show that MH-DQN outperforms other DRL algorithms in single server in terms of convergence speed and system utility. Then, we validate that cache decision models incorporating the personalized layers perform better in heterogeneous environments compared to fully-federated and non-federated methods.

\section{System Model and Problem Formulation}

\subsection{Network Model}
\begin{figure}[!t]
	\centering
	\includegraphics[width=1\linewidth]{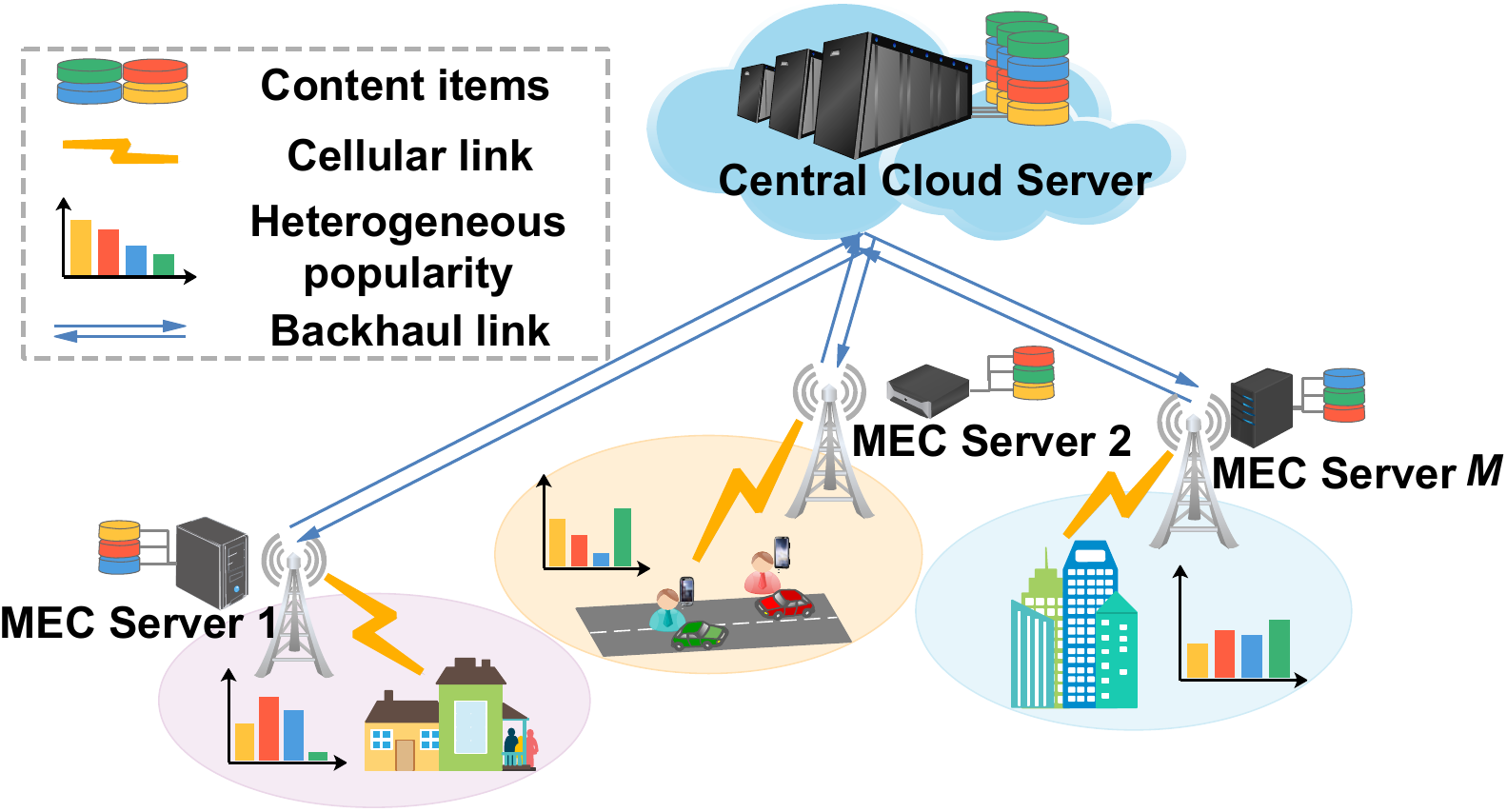}
	\caption{System model.}
	\label{SystemModel}
 \vspace{-0.2in}
\end{figure}

We consider a system comprising a central cloud server (CCS) and a network of mobile edge caching (MEC) servers, denoted by $\mathcal{M} = \{1,2,\dots,M\}$. Each MEC server is connected to a corresponding base station and provides content caching service to users. The system manages a total of $C$ popular content items, represented as $\mathcal{C}=\{1,2,\dots,C\}$, which can be requested by users. Each content $c\in \mathcal{C}$ is characterized by a tuple $(\eta_c, \phi_c)$, where $\eta_c \in \mathbb{R}^+$ and $\phi_c \in \mathbb{R}^+$ represent the data size and the downloading payment of content $c$, respectively. The set of users is defined as $\mathcal{U}=\{1,2,\dots,U\}$, with each subset $\mathcal{U}_m \subseteq \mathcal{U}$ representing the users within the coverage area of MEC server $m$. To simplify the model, we assume that the coverage areas of different MEC servers do not overlap. The system model is depicted in Fig.~\ref{SystemModel}.

\subsection{Proactive Caching Model}
The system operates over finite time slot set $\mathcal{T}= \{1,2,\dots,T\}$ and each slot contains the following stages:

$\bullet$ \textit{Content Placement.} At the beginning of time slot $t$, the content cached at MEC server $m$ is updated according to the caching policy $\boldsymbol{a}_m^t = \{a_{m,1}^t, a_{m,2}^t,\dots,a_{m,C}^t\}$, where $a_{m,c}^t = 1$ if the MEC server $m$ caches content $c$ at time slot $t$, otherwise $a_{m,c}^t =0$. Let $N_m \in \mathbb{R}^+$ be the storage capacity of MEC server $m$. Thus, the caching policy $\boldsymbol{a}_m^t$ must adhere to the following condition
\begin{equation}
\begin{array}{llll}
	\sum\nolimits_{c=1}^{C} a_{m,c}^t\eta_c \leq N_m,\  \forall m \in \mathcal{M}, \ \forall t\in\mathcal{T}.
 \end{array}
\end{equation}

$\bullet$ \textit{User Request.} After content placement, user $u\in\mathcal{U}_m$ sends a request vector $\boldsymbol{d}_u^t = \{ d_{u,1}^t, d_{u,2}^t, \dots, d_{u,C}^t\}$ to MEC server $m$, where $d_{u,c}^t = 1$ means the user $u$ request the content $c$ at time slot $t$, otherwise $d_{u,c}^t =0$. Thus, for each MEC server $m$, the revived request vector is defined as $\boldsymbol{d}_m^t = \{ d_{m,1}^t, d_{m,2}^t, \dots, d_{m,C}^t\}$, where $d_{m,c}^t = \sum_{u\in U_m}d_{u,c}^t$, representing the content popularity $P_{c,m}$ in the service area of $m$. In the proactive caching model, content popularity $P_{c,m}$ is generally assumed to be \textbf{unknown} and needs to be learned. Besides, considering that user demographics and behavior could be influenced by their geographical location, we point out that the popularity of the same content varies across different MEC servers, exhibiting \textbf{heterogeneity}, i.e., $P_{c,m}\neq P_{c,n}$ for $m\neq n$.  For instance, some works model the content popularity distribution as the Mandelbrot-Zipf (MZipf) distribution~\cite{LeeM}, where the popularity of content $c$ at server $m$ is expressed as
\begin{equation}
	\label{mzipf}
	P_{c,m} = \frac{(\kappa_c + q_m)^{-k_{m}}}{\sum_{i=1}^{C}(\kappa_i+q_m)^{-k_{m}}},
\end{equation}
where $\kappa_c$ represents the popularity rank of content $c$ and $q_m$ and $k_m$ denote the plateau factor and Zipf factor. The heterogeneity of content popularity is reflected by the condition $q_m \neq q_n, k_m \neq k_n,$ for $m,n\in\mathcal{M}, m\neq n$. 

$\bullet$ \textit{Cache Service.} At the end of each time slot, user requests are fulfilled either by edge delivery (i.e., a cache hit) or via remote delivery from the CCS over the backhaul links (i.e., a cache miss).

\subsection{System Utility Model}
In this work, the system utility is composed of two parts:
\subsubsection{Replacement Cost}
From the perspective of the MEC servers, downloading content from the CCS can be costly. Thus, the large replacement cost incurred by frequent content fetching should be avoided. Denote the downloading payment of content $c$ is $\phi_c$, then the replacement cost of MEC server $m$ at time slot $t$ can be expressed as
\begin{equation}
    E_m^t = \sum\nolimits_{c=1}^{C} \mathbb{I}(a_{m,c}^t, a_{m,c}^{t-1})\phi_c,
\end{equation}
where $\mathbb{I}(i, j) \in \{0,1\}$ is an indicator function that returns $1$ if  $i=1$ and $j=0$, and $0$ otherwise. 
\subsubsection{Cache Hit Ratio}
Regarding the users, a cache hit allows immediate edge delivery via cellular links, while a cache miss requires fetching the content from remote CCS via backhaul links, leading to larger delays. Therefore, the Cache Hit Ratio (CHR), as the metric used to evaluate the user QoE, is  another factor of our system utility. Specifically, the instantaneous CHR can be calculated as
\begin{equation}
	H_m^t = \frac{\sum_{u\in\mathcal{U}_m}\sum_{c=1}^{C}a_{m,c}^td_{u,c}^t}{\sum_{u\in\mathcal{U}_m}\sum_{c=1}^{C}d_{u,c}^t}.
\end{equation}

\subsection{Problem Formulation}
In this paper, we aim to improve the system utility over a time horizon of $T$ by simultaneously maximizing the CHR and minimizing the replacement cost, while considering the limited storage capacity. The optimization problem is formulated as
\begin{equation} 
	\label{opt}
	\max _{\{\boldsymbol{A}\}} \ {\sum_{t=1}^{T}\sum_{m=1}^{M}\omega_1 H_m^t-\omega_2 E_m^t},\\
\end{equation}
\begin{equation}
	\text{s.t.}	
	\begin{cases}
		\text{C1:}\ \sum_{c=1}^{C} a_{m,c}^t\eta_c \leq N_m,\  \forall m \in \mathcal{M}, \ \forall t\in\mathcal{T},\\
		\text{C2:}\ a_{m,c}^t\in\{0,1\}, \ \forall m\in\mathcal{M},\  \forall t \in\mathcal{T}, 
	\end{cases}
 \notag
\end{equation}
where $\boldsymbol{A} = \{\boldsymbol{a}_1^t, \boldsymbol{a}_2^t, \dots, \boldsymbol{a}_M^t,~ \forall t\in \mathcal{T}\}$ represents the caching decision variables of MEC servers, $\omega_1$ and $\omega_2$ are the weighted parameters for two utility metrics. 
The constraint (C1) ensures that the sum of the content size will not exceed the storage capacity. The constraint (C2) ensures that our caching actions are discrete and indivisible. The challenges in solving this problem using FDRL lie in three aspects: 
1) \textit{High-dimensional action space}: The demand caching policy needs to explore a vast combination of possible discrete actions. 
2) \textit{Unknown popularity}: The proactive caching scheme requires the caching policy to be determined solely based on historical user requests and feedback. 
3) \textit{Heterogeneous popularity}: This leads to a trade-off between the usage of global knowledge and the local model. Relying solely on limited local observation leads to insufficient learning and slow convergence, while naively using global information from other MEC servers may overlook the local bias. 


\section{Algorithm Design}
In this section, we first propose a multi-head structure to improve the existing DQN algorithm, enabling a single MEC server to handle unknown content popularity and an exploding action space. Subsequently, we construct a personalized federated learning architecture that allows multiple MEC servers to enhance learning performance in a privacy-preserving way while simultaneously accommodating local heterogeneity in content popularity.
\subsection{Multi-head Deep Q-Network for Single MEC server}
To implement the DRL method, we first formulate the proactive caching process as a Markov Decision Process (MDP), which can be defined by a 4-tuple $(\mathbb{S}, \mathbb{A}, \mathbb{P}, \mathbb{R})$, where 
$\mathbb{S}$ represents the set of system states, 
$\mathbb{A}$ represents the set of actions, 
$\mathbb{P}_a(s, s')$ is the set of probability that action $a$ in state $s$ will lead to state $s'$,
$\mathbb{R}_a(s, s')$ consists of reward values after transitioning from state $s$ to state $s'$, due to action $a$. For each MEC server $m$, the detailed state, action space, and reward function are summarized as follows.
\subsubsection{State Space}
The state is the system information observed by the MEC server $m$ at the beginning of each time slot,  described as $\boldsymbol{s}_{t}^{m}= [\boldsymbol{\bar{d}}_{m}^t, \boldsymbol{a}_m^{t-1}] \in \mathbb{S} $. Here $\boldsymbol{a}_m^{t-1}$ is the cache placement decision made by the last time slot, indicating the current cache status. $\boldsymbol{\bar{d}}_{m}^t$ is the \textit{historical} observation of the content request, defined as 
\begin{equation}
   \boldsymbol{\bar{d}}_{m}^t = \frac{\sum_{k=1}^{w-1}\beta^k\cdot\boldsymbol{d}_m^{(t-k)}}{\sum_{k=1}^{w-1}\beta^k}.
\end{equation}
We adopt an exponentially weighted moving average (EWMA) approach to model content popularity, with window size as $w$ and decay factor $0 <\beta<1$, which assigns exponentially decreasing weights over time to past observations, thereby allowing the model to adapt more rapidly to temporal variations and capture the latest popularity trends more effectively.

\subsubsection{Action Space}
The MEC server  $m$ determines which content should be cached for the next time slot. Thus, the action vector is described as $\boldsymbol{a}_m^t = [a_{m,1}^t, a_{m,2}^t, \dots,a_{m,C}^t]$. Since each element $a_{m,c}^t$ can take on two possible values, $0$ or $1$, the action space size for each MEC server is $2^C$, growing exponentially with the number of contents $C$.

\subsubsection{Reward Function}
Given the state-action pair, the immediate reward obtained from the environment is defined as
\begin{equation}
    \label{reward}
	r_m^t = \omega_1 H_m^t-\omega_2 E_m^t -\omega_3 \varrho,
\end{equation}
where the first two items are related to the objective function, which aims to maximize the CHR and minimize the replacement cost. The last term $\varrho$ is the penalty cost  incurred when the caching action violates the storage constraint (C2).

After formulating our problem as an MDP, we employ RL-based methods to optimize the caching actions by evaluating the long-term rewards associated with state-action pairs. Traditional RL techniques, such as Q-learning, maintain a tabular representation of the state-action value function $Q(s,a)$, with each entry storing the expected return for taking action $a$ in state $s$. However, such tables become infeasible when dealing with high-dimensional state and action spaces common in caching problems. 
The DQN adopts the deep neural networks, known as the Q-networks, as the non-linear function approximators to take the current state as input and output Q-values for all possible actions, subsequently selecting the action associated with the highest Q-value as the optimal choice.
While DQN can effectively handle large state spaces, it still struggles with exponentially growing discrete action spaces as the number of contents $C$ increases. 


\begin{figure}[!t]
	\centering
	\includegraphics[width=0.98\linewidth]{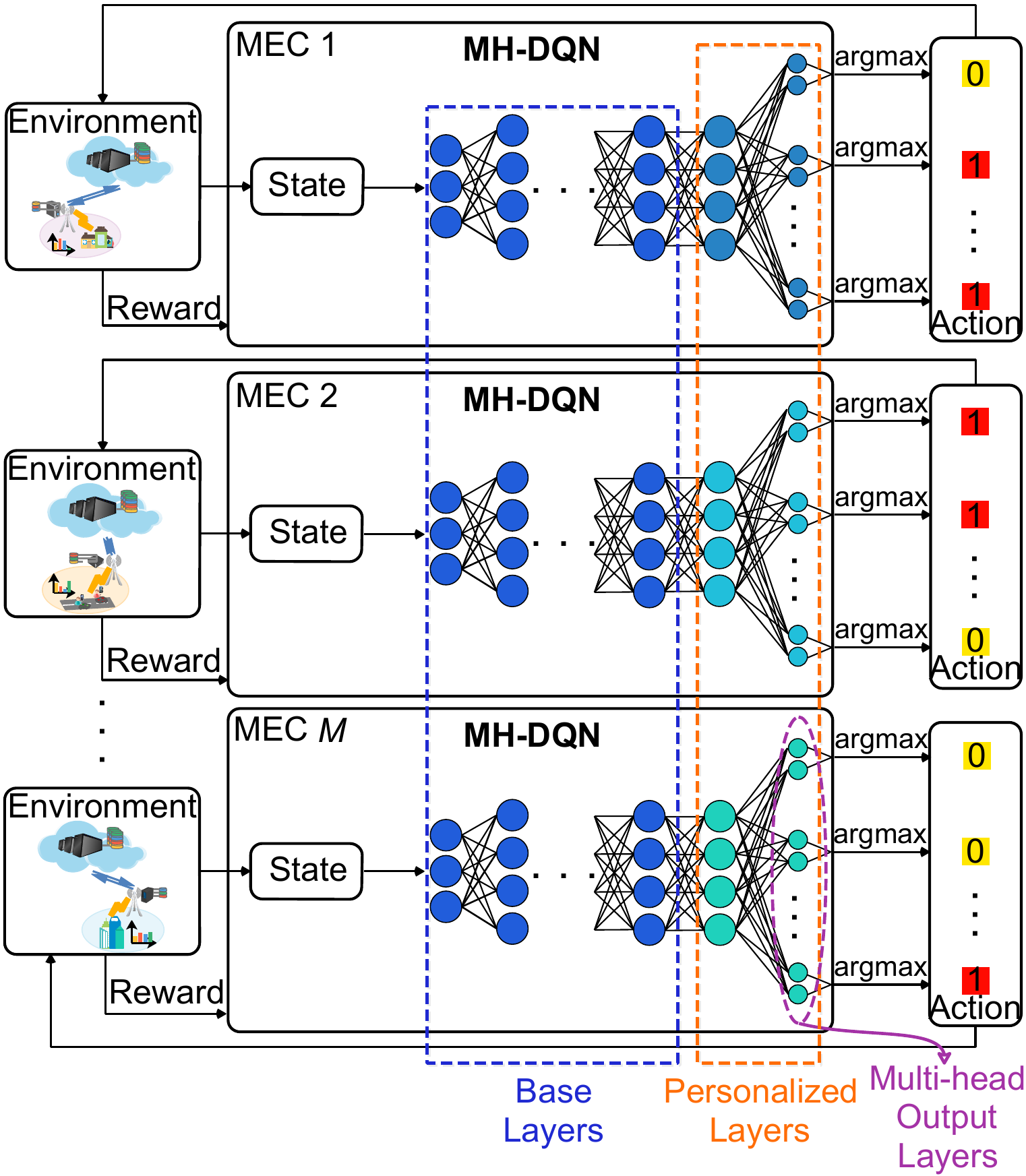}
	\caption{Overview of proposed PF-DRL-Ca framework.}
	\label{framework}
	\vspace{-0.2in}
\end{figure}
To tackle the curse of dimensionality arising from the exponential growth of the discrete action space, we propose a modified DRL architecture called the multi-head DQN (MH-DQN). Instead of the conventional single output layer in the Q-network that estimates Q-values for all $2^C$ possible cache actions, we replace it with $C$ parallel output layers, referred to as ``\textit{heads}''. Each of these heads is responsible for outputting the Q-value estimates corresponding to the two sub-actions (cache or not cache) for one individual content. This multi-head design effectively decomposes the original exponentially large joint action space into $C$ independent binary action spaces, one for each content. As a result, the total output size now scales linearly with the number of contents $C$, instead of exponentially,
leading to a change in the number of output spaces from $2^C$ to $2C$ in our caching decision-making problem.

The details of the multi-head structure are shown in the purple oval box in Fig.~\ref{framework}. The state vector is taken as input to output the sub-Q-value functions for each action dimension $c\in\{1,\dots,C\}$. In particular, the sub-Q-value of sub-action $a_{m,c}\in\{0,1\}$ in the $c$-th action dimension is defined as $Q_{m,c}(\boldsymbol{s}_m^t, a_{m,c})$. For each action dimension $c$, the sub-action with the highest Q-value is chosen as the dimension's representative, that is 
\begin{equation} \label{action}
\begin{array} {llll}
a_{m,c}^t = \arg\max_{a_{m,c}\in\{0,1\}} Q_{m,c} (\boldsymbol{s}_m^t,a_{m,c}). 
\end{array}
\end{equation}
The final joint action $\boldsymbol{a}_m^t$ is generated by combining each dimension action, i.e., $\boldsymbol{a}_m^t = \{ a_{m,c}^t|c = 1,\dots,C\}$.

In the MH-DQN, there exists another target network $Q^{-}_{m,c} (\cdot)$ that has the same $C$ parallel heads with the Q-network, 
which is used to compute the approximated Q-values for updating the Q-network, defined as
\begin{align} \label{target}
	&y_{m,c}^t = r_{m}^t + \gamma \max_{a_{m,c}\in\{0,1\}} Q^{-}_{m,c}(\boldsymbol{s}_m^{t+1},a_{m,c}),
\end{align}
where $\gamma$ is the discounted factor.
During the training process, the MEC server $m$ chooses a mini-batch of transitions $(\boldsymbol{s}_m^t,\boldsymbol{a}_m^t,r_m^t, \boldsymbol{s}_m^{t+1}) $ from the replay buffer as the training samples. We update the Q-network parameters  by minimizing the loss function, which can be given as
\begin{equation}\label{loss}
\mathcal{L}(\theta^t_m) = \mathbb{E}_{(\boldsymbol{s}_m^t, \boldsymbol{a}_m^t, r_m^t, \boldsymbol{s}_m^{t+1})} [\frac{1}{C}\sum_{c=1}^C(y_{m,c}^t - Q_{m,c}(\boldsymbol{s}_m^t, a_{m,c}^t))^2]. 
\end{equation}
The parameters $\theta_m$ of the Q-network are updated using gradient descent based on the loss calculated above, while the parameters $\theta_m^{-}$ of the target network are softly updated to slowly align with the Q-network, ensuring stability in the learning process. The primary steps of the MH-DQN algorithm are illustrated in Algorithm~\ref{Alg1}.

\begin{algorithm}[!t]
	\caption{MH-DQN Algorithm for MEC Server $m$}
	\label{Alg1}
	\begin{algorithmic}[1] 
		\State Initialize replay buffer $\mathfrak{B}_m$, learning rate $\xi \in [0,1]$, update coefficient $\tau \in [0,1]$;
		\State Initialize Q-network with random parameters $\theta_m$ and target network with parameters $\theta_m^- = \theta_m$;
		\For{episode $ e = 1,\dots,E$}
		\State Initialize state $\boldsymbol{s}_m^0$ randomly;
		\For{$t = 0, 1, 2, \dots, T$}
    \State With probability $\epsilon$ select  $\boldsymbol{a}_m^t$ randomly; Otherwise select $\boldsymbol{a}_m^t = \{ a_{m,c}^t|c = 1,\dots,C\}$ by Eq.~\eqref{action};
    \State Cache content based on $\boldsymbol{a}_m^t$, observe next state $\boldsymbol{s}_m^{t+1}$ and reward $r_m^t$ according to Eq.~(\ref{reward});
    \State Store transition $(\boldsymbol{s}_m^t, \boldsymbol{a}_m^t, r_m^t, \boldsymbol{s}_m^{t+1})$;
    \State Sample a mini-batch of transitions randomly from $\mathfrak{B}_m$ and calculate the loss function in Eq.~(\ref{loss});
    \State Update Q-Network by:
    \Statex $\theta^{t+1}_m = \theta^{t}_m - \xi\nabla_{\theta^t_m}(\mathcal{L}(\theta^t_m))$;
    \State Update target network by:
    \Statex $\bar{\theta_{m}^{t+1}} = \tau \theta_m^{t+1} + (1-\tau) \bar{\theta_{m}^{t}}$;
    \EndFor
    \EndFor
\end{algorithmic}  
\vspace{-0.2in}
\end{algorithm}  

\subsection{Personalized Federated DRL for Caching}
Based on the proposed MH-DQN architecture that enables efficient learning of caching policies over large content size, we next consider how to coordinate and improve the decision models across MEC servers through federated learning. Notably, FDRL frameworks have recently shown promising results in this regard. The core idea behind FDRL is to aggregate locally trained DRL models from individual MEC servers into a more robust global model, without requiring raw user data to be shared, thereby preserving privacy.

However, simply replacing each MEC server's local model with the globally aggregated model can lead to an undesirable homogenization of the caching strategies across all MEC servers \cite{pfdrl,pfdrl2}. This naive approach overlooks the crucial fact that different MEC servers may serve diverse user populations with heterogeneous content preferences and access patterns. Consequently, enforcing a \textit{one-size-fits-all} global policy fails to adapt to such localized heterogeneity, leading to suboptimal performance. To address this limitation, we propose a Personalized FDRL Caching (PF-DRL-Ca) architecture that supports both layer-wise and MEC server-wise personalization. 

We first propose a layer-wise aggregation scheme for the PF-DRL-Ca architecture. Specifically, assuming there are a total of $L$ layers in the Q-network (as well as the target network) proposed in Algorithm~\ref{Alg1}, we manually divide the neural network layers $l \in \{1,\dots,L\}$ into base layer set $L_B$ and personalized layer set $L_P$, as illustrated in Fig.~\ref{framework}. The base layers, located closer to the model input, are responsible for extracting relevant features from the state vector representing the user request patterns. These lower layers learn to enhance the state perception and encode generalizable knowledge shared across all MEC servers. In contrast, the layers closer to the output, tasked with evaluating and outputting the caching actions, are designated as personalized layers. This segmentation allows the base layers to learn generalizable features shared across all MEC servers, while the personalized layers specialize in learning localized caching policies tailored to the unique request characteristics of the user populations served by each individual MEC server. During the federated learning process, only the weights of the base layers are aggregated across all MEC servers to form a global shared representation. The personalized layers, on the other hand, retain their locally trained weights, enabling each MEC server to maintain a personalized decision-making component adapted to its specific environment. The effects of this layer segmentation strategy on learning performance are thoroughly explored in the experimental section.

After segmenting the base and personalized layers, we proceed with the federated training process as shown in Algorithm~\ref{Alg2}. Both the local and global models are initialized with the same MH-DQN architecture. Each training round begins with local updates to the Q-network and target network at each MEC server, following the procedure specified in Algorithm~\ref{Alg1}. At the end of each episode, the MEC servers upload the parameters of the base layers in $L_B$ to the CCS, along with the total number of requests $D_m(e)$ received during that episode (Line 6). Here, $D_m(e)$ serves as an indicator of the contribution made by each MEC server $m$ during the current episode $e$.
MEC servers that receive more user requests will obtain more feedback samples, enabling more sufficient training of their local models. Therefore, these MEC servers should have a larger influence on the global model aggregation. Consequently, the CCS aggregates the parameters of the base layers using a weighted sum scheme, prioritizing MEC servers with larger $D_m(e)$ values. Specifically, the aggregated base layer parameters $\theta_G(l)$ for layer $l \in L_B$ are computed as: $\theta_G(l) = \sum_{m=1}^M \frac{D_m(e)}{\sum_{m=1}^M D_m(e)} \cdot \theta_m(l)$, where $\theta_m(l)$ represents the locally trained base layer parameters at MEC server $m$. This weighted averaging assigns higher importance to MEC servers that have processed more requests, thereby allowing them to contribute more significantly to the global base representation.

\begin{algorithm}[!t]
    \caption{PF-DRL-Ca Algorithm}
    \label{Alg2}
    \begin{algorithmic}[1] 
        \State Initialize base layer set $L_B$, personalized layer set $L_P$;
        \State Initialize MH-DQN models for all MEC servers and CCS;
        
        \For{episode $e = 1,\dots,E$}
            \For{each MEC server $m \in M$}
                \State Train MH-DQN models $\theta_m$, $\bar{\theta}_m$ using Alg.~\ref{Alg1};
                 \State $D_m(e) = \sum_{t = 1}^{T} \sum_{c=1}^C d_{m,c}^t$;\Comment{\textit{Calculate weight}}
                \State Send $D_m(e)$ and $\theta_m(l)$ to CCS for each $l \in L_B$;
            \EndFor
            
            \For{$l \in L_B$} \Comment{\textit{Aggregation step}}
                \State CCS aggregates base layers $l$:
                \State $\theta_G(l) = \sum_{m=1}^M \frac{D_m(e)}{\sum_{m=1}^M D_m(e)} \cdot \theta_m(l)$;
            \EndFor
            
            \State CCS broadcasts $\theta_G(l)$ for all $l \in L_B$ to MEC servers;
            
            \For{each MEC server $m \in M$}
                \For{each layer $l \in L_B$}
                    \State $\theta_m(l) = \bar{\theta}_m(l) = \theta_G(l)$; \Comment{\textit{Sync base layers}}
                \EndFor
                \EndFor
            \EndFor
    \end{algorithmic}
\end{algorithm}

\section{Performance Evaluation}

\subsection{Experimental Settings}
We consider there are five MEC servers within the system, and the system includes two types of heterogeneity: content popularity and user distribution.
We model heterogeneous content popularity by setting different MZipf plateau factors and Zipf factors in Eq.~\eqref{mzipf}~\cite{WangC}.
For the user distribution, we assume that the number of users within the coverage area of each MEC server is not the same.
The DNN structures of proposed algorithms are implemented by the deep learning library PyTorch~1.8. 
The detailed default experimental parameters are listed in Table~\ref{para}.

\begin{table}[!t]
	\begin{center}
		\caption{Simulation Parameters}
		\label{para}
		\begin{tabular}{cc} 
			\toprule 
			\textbf{Parameter} & \textbf{Value} \\
			\midrule 
			Number of MEC servers $M$ & 5 \\
			Plateau factors of MZipf $q_m$ & [100, 200, 90, 40, 80] \\
			Zipf factors of MZipf $k_m$ & [0.60, 0.60, 0.75, 0.90, 0.90] \\
			Data size of contents $\eta_c$ & 1$\sim$8 GB \\
			Downloading payment of contents $\phi_c$ & 0.05$\sim$0.5 HKD \\
			Number of DNN layers & 6 \\
			Number of neurons per hidden layer & 128 \\
			Learning rate $\xi$ & 0.002 \\
			Update coefficient $\tau$ & 0.005 \\
			Discounted factor $\gamma$ & 0.99 \\
			\bottomrule 
		\end{tabular}
	\end{center}
\vspace{-0.2in}
\end{table}

\subsection{Performance Evaluation of MH-DQN Algorithm}
In this subsection, we set the number of contents to $40$ and evaluate the effectiveness of the MH-DQN algorithm for a single MEC server. 
Since traditional DQN cannot handle such a large action space, we use continuous algorithm PPO~\cite{WangY} with a sampling process in the same environment for comparison. 

\begin{figure}[!t]
	\centering
	\includegraphics[width=1\linewidth]{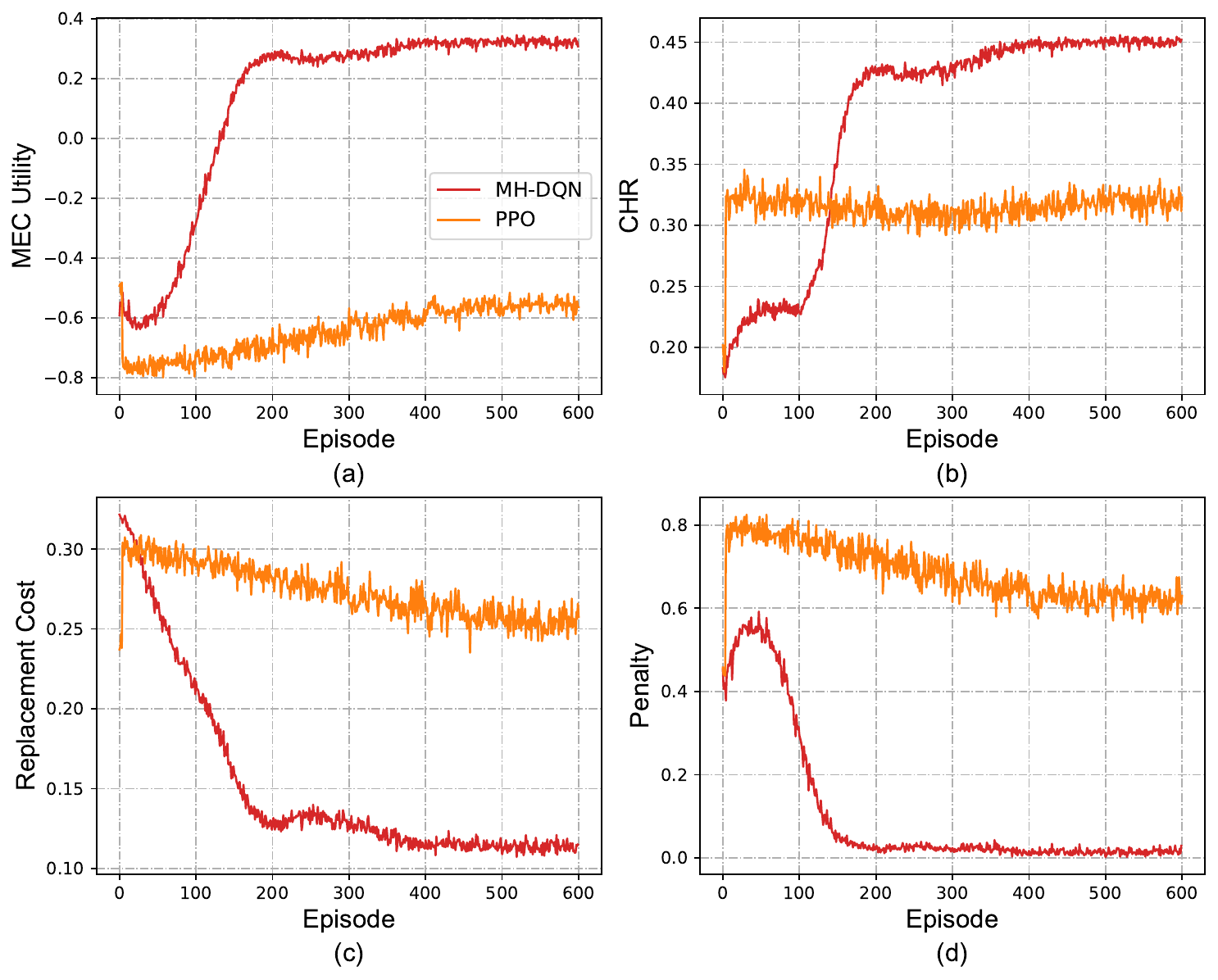}
	\caption{Performance comparison of modified DQN and PPO algorithm.}
	\label{ppo}
	\vspace{-0.1in}
\end{figure}

The results across multiple metrics consistently show the superiority of our proposed MH-DQN approach over the baseline PPO. 
In Fig.~\ref{ppo}(a), which displays MEC utility, MH-DQN demonstrates a significant improvement, stabilizing at approximately 0.35 after an initial rise. In contrast, PPO shows greater fluctuation and achieves lower utility values consistently. 
Fig.~\ref{ppo}(b) shows the CHR, where MH-DQN reaches a stable peak close to 0.45, whereas PPO exhibits lower and more variable CHR values. 
Fig.~\ref{ppo}(c) illustrates replacement costs, where MH-DQN's costs decrease quickly and remain low, indicating efficient resource management. On the other hand, PPO's costs are higher and display more fluctuations. 
In Fig.~\ref{ppo}(d), concerning penalties, MH-DQN quickly reduces penalties to nearly zero, showing its effectiveness in minimizing operational drawbacks. PPO, however, maintains higher levels of penalties throughout the episodes.

We also observe that PPO exhibits higher policy entropy fluctuations, suggesting issues with premature convergence or unstable learning dynamics arising from directly optimizing the high-dimensional caching policy over the exponentially large action space. Our MH-DQN framework solves this through the factorized multi-head representation, which reduces policy sensitivity to parameter updates, enabling smoother convergence. Consequently, MH-DQN learns more coherent and stable caching policies, contributing to its superior overall performance.

\subsection{Performance Evaluation of PF-DRL-Ca Algorithm}
\begin{figure}[!t]

	\centering
	\includegraphics[width=0.68\linewidth]{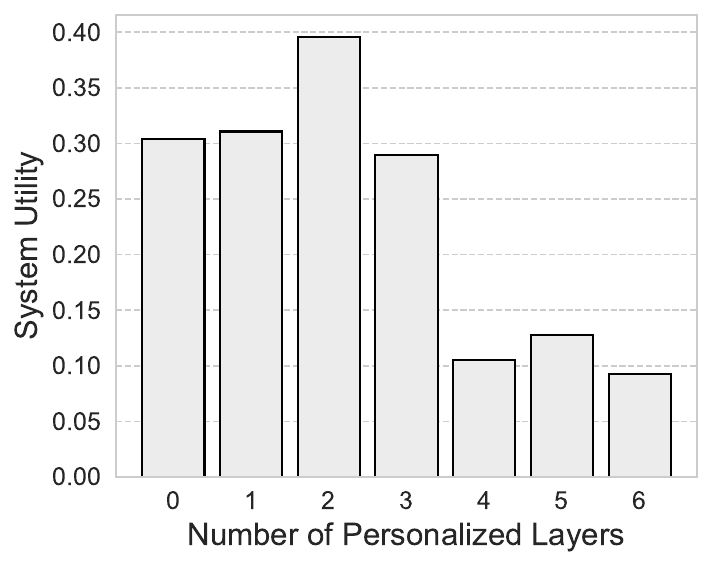}
	\caption{Performance comparison with different numbers of personalized layers.}
	\label{NumOfPer}
 \vspace{-0.1in}
\end{figure}
\begin{figure}[!t]
	\centering
	\includegraphics[width=0.68\linewidth]{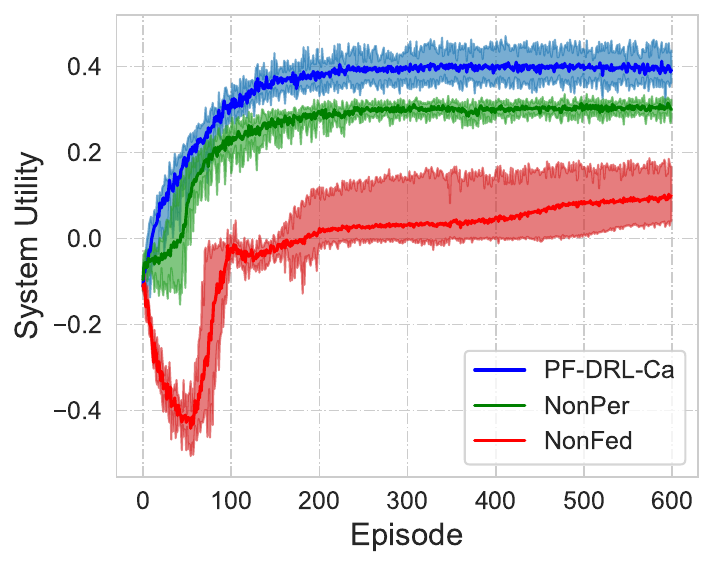}
	\caption{Convergence comparison of three DRL-based algorithms.}
	\label{converagence}
\vspace{-0.1in}
\end{figure}
We evaluate and compare the performance of the PF-DRL-Ca algorithm with two learning-based algorithms and three non-learning caching approaches, which are presented as follows.
\romannumeral1) \textit{NonPer}: A conventional FDRL framework without the personalized layers. It can also be viewed as the number of personalized layers is $0$;
\romannumeral2) \textit{NonFed}: A conventional DRL framework without the FL paradigm. It can also be viewed as the number of personalized layers equals $6$;
\romannumeral3) \textit{LRU}~\cite{HuangQ}: A non-learning approach in which the least recently used contents will be replaced;
\romannumeral4) \textit{LFU}~\cite{HuangQ}: A non-learning approach in which the least frequently used contents will be replaced;
\romannumeral5) \textit{Random}: A non-learning approach in which content caching decisions are generated randomly.
%
%
%
%
%
%


\subsubsection{Effect of the Number of Personal Layers}

Fig.~\ref{NumOfPer} shows the impact of varying personalized layer counts on the system utility achieved by our PF-DRL-Ca algorithm. The utility initially increases as personalized layers are added from zero up to an optimal point of two personalized layers, after which adding more layers provides no further benefit. 
Fig.~\ref{NumOfPer} also shows that employing two personalized layers yields a significant 30\% utility gain over NonPer and a remarkable 330\% improvement compared to NonFed. This result highlights the importance of balancing shared global representations with localized personalization. 
Without personalized layers, the servers are unable to handle the heterogeneity in their environments. 
Conversely, with all layers personalized, the algorithm fails to capture commonalities across servers, hindering knowledge sharing. 
The optimal configuration of four base layers and two personalized layers allows base layers to extract generalizable features from aggregated data, while personalized layers adapt to each server's unique user preferences. 
Therefore, we set this layer configuration for the following experiments.

\subsubsection{Convergence Performance} As shown in Fig.~\ref{converagence}, our proposed PF-DRL-Ca method outperforms the two baselines, NonPer and NonFed, in terms of both final converged utility and convergence speed. PF-DRL-Ca achieves the highest system utility while maintaining a relatively narrow shaded region, suggesting that it effectively balances overall performance and local heterogeneity across MEC servers. In contrast, the NonFed approach, which lacks any federated collaboration, exhibits the lowest utility and the widest shaded region, indicating significant performance disparities among MEC servers. This is expected as the absence of knowledge sharing leads to incomplete training for each individual server, resulting in suboptimal and highly heterogeneous caching strategies. The NonPer method, while incorporating federated aggregation through the CCS, still underperforms PF-DRL-Ca. Despite achieving a narrower shaded region compared to NonFed, the lack of personalization in NonPer leads to a \textit{one-size-fits-all} global policy that fails to adapt to local user preferences, resulting in a lower overall utility.  By separating the neural network into shared base layers and personalized  layers, PF-DRL-Ca effectively captures both global patterns and local heterogeneities. The weighted aggregation of base layers prioritizes contributions from MEC servers with more user requests, further enhancing the global representation. Simultaneously, the personalized head layers enable each server to specialize in catering to its unique user population, leading to optimized caching strategies that strike the right balance between global coordination and local adaptation.

\begin{figure}[!t]
	\centering
	\includegraphics[width=0.68\linewidth]{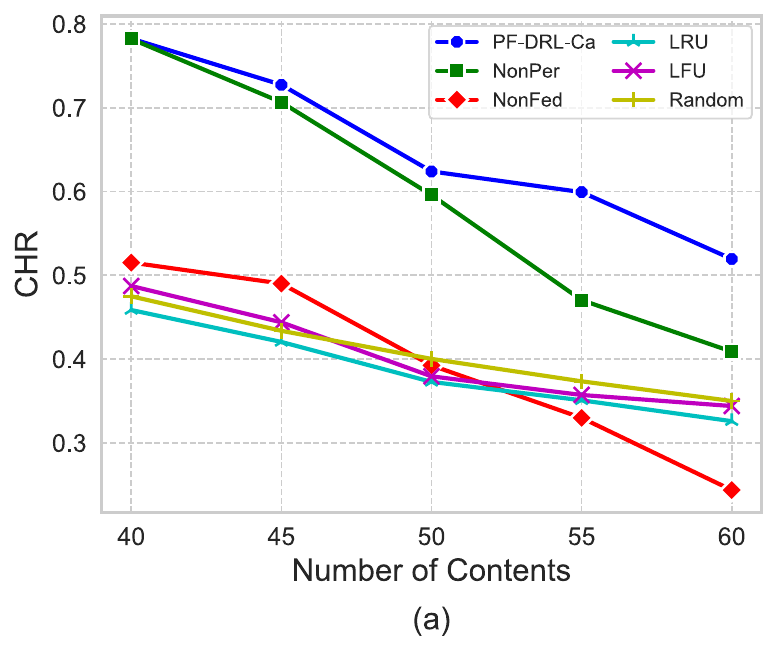}
 \includegraphics[width=0.68\linewidth]{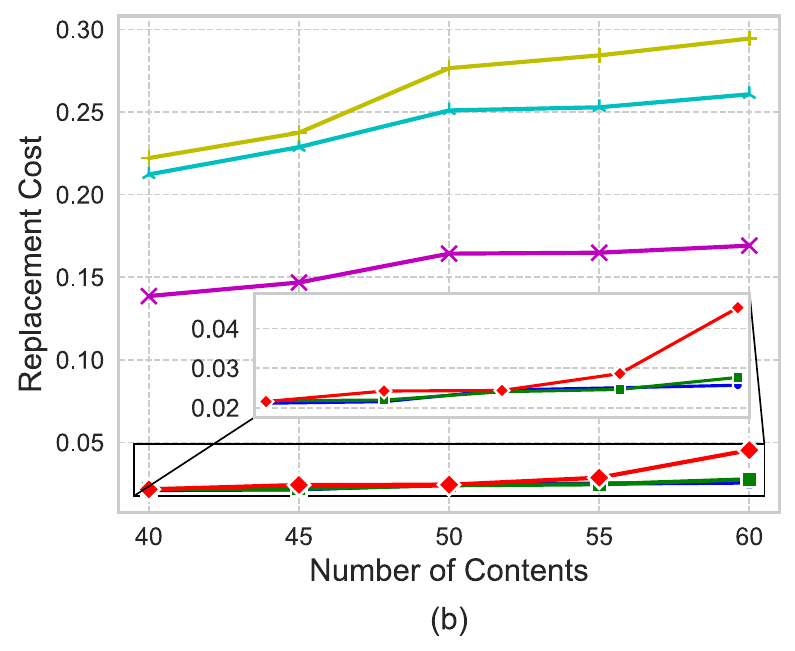}
	\caption{Performance comparison with different numbers of contents in terms of (a) CHR; (b) replacement cost.}
	\label{ContentsNum}
	\vspace{-0.1in}
\end{figure}

\subsubsection{Effect of the Number of Contents}

Fig.~\ref{ContentsNum} presents a performance comparison of six caching algorithms as the number of contents varies. Fig.~\ref{ContentsNum}(a) depicts the average CHR for different algorithms as the number of contents increases from 40 to 60. We observe a declining trend in the CHR for all algorithms as more contents are introduced. This behavior can be attributed to the fixed cache capacity limitation. With a larger content catalog, the same cache size can hold a smaller fraction of the total contents. This leads to a lower probability of requested contents being available in the cache, thereby reducing the overall CHR. Despite this general decline, our proposed PF-DRL-Ca algorithm consistently outperforms all baselines, maintaining the highest average CHR across the entire range of content numbers. Moreover, the performance gap between PF-DRL-Ca and the best alternative baseline widens as the number of contents grows, highlighting its effectiveness in optimizing cache utilization under increasing decision space complexity. Fig.~\ref{ContentsNum}(b) reveals that all learning-based methods, including PF-DRL-Ca, NonPer, and NonFed, exhibit comparable and significantly lower average replacement costs than non-learning approaches. This advantage can be attributed to the ability of DRL models to learn and adapt to user request patterns, enabling more informed caching decisions from the outset and minimizing the need for frequent cache replacements.

\begin{figure}[!t]

		\centering
		\includegraphics[width=0.68\linewidth]{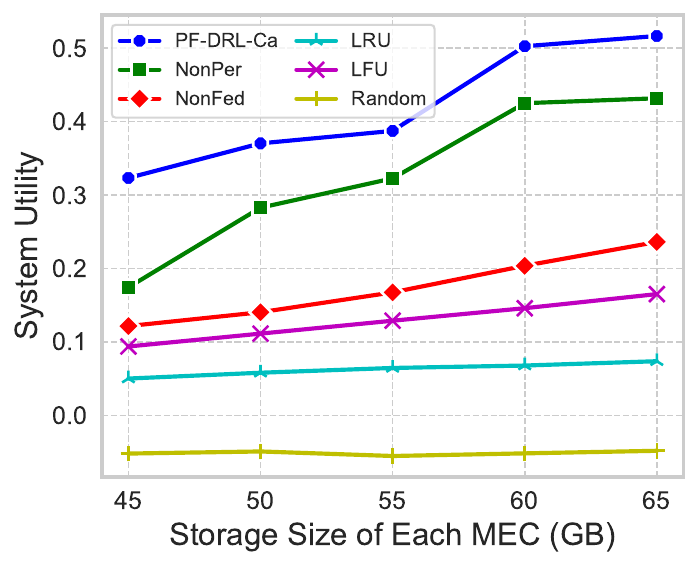}
		\caption{Performance comparison with different storage sizes.}
		\label{StorageSize}
  \vspace{-0.1in}
  \end{figure}

	\begin{figure}[t]
		\centering
		\includegraphics[width=0.68\linewidth]{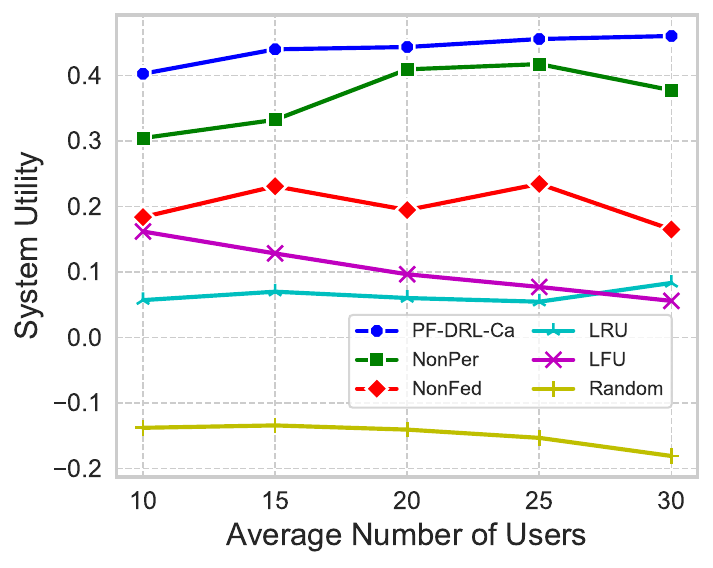}
		\caption{Performance comparison with different average numbers of users.}
		\label{UsersNum}
  \vspace{-0.1in}
	\end{figure}


\subsubsection{Effect of the Cache Size and User Dynamics}

Fig.~\ref{StorageSize} presents a performance comparison of different caching algorithms under varying storage sizes for each MEC server, ranging from 45 GB to 65 GB. As expected, the system utility of all algorithms increases with larger storage sizes. This trend is intuitive because a larger cache capacity allows more contents to be stored, increasing the likelihood of serving user requests from the cache and improving overall utility. Our proposed PF-DRL-Ca algorithm consistently outperforms all baselines across the entire range of storage sizes, achieving the highest system utility. Notably, the performance gap between PF-DRL-Ca and the other algorithms widens as the storage size increases, highlighting its ability to leverage the additional cache capacity through intelligent caching decisions.

Fig.~\ref{UsersNum} examines the performance of the same algorithms under different average numbers of users per MEC server. In this scenario, we observe an interesting trend: the system utility of PF-DRL-Ca and the NonPer baseline increases as the average number of users grows, while the utility of the NonFed and non-learning algorithms, like LFU and Random, decreases. This contrasting behavior can be attributed to the ability of learning-based algorithms, particularly PF-DRL-Ca, to adapt and optimize caching strategies based on user request patterns. With a larger user base, these algorithms can leverage more diverse request data to improve their caching policies, leading to better cache utilization and higher system utility. In contrast, the non-learning algorithms, such as LFU and Random, cannot adapt to changing user dynamics, resulting in suboptimal caching decisions and a decline in performance as the user base grows more diverse and complex. 

These experimental results demonstrate the robustness and effectiveness of our proposed PF-DRL-Ca algorithm in handling varying cache sizes and user dynamics, reinforcing its suitability for intelligent caching in dynamic edge computing environments.


\section{Conclusion}
In this paper, we propose the PF-DRL-Ca approach for multi-server proactive caching,  effectively tackling three key challenges: immense action space, unknown content popularity, and heterogeneous content requests. We utilize a multi-head structure to reshape the DQN's output layer, making the action output space scale linearly with the number of contents. 
We further propose a personalized federated training architecture. Uniquely, each MEC server uploads only the base layers, which are essential for environmental perception, to the cloud for aggregation, while retaining the strategy-oriented layers as personalized layers. This strikes a balance between leveraging the benefits of collaborative knowledge sharing and allowing individual servers to adapt to their local environments.

\bibliographystyle{IEEEtran}
\bibliography{IEEEtran,wiopt}
\end{document}